\documentclass[12pt]{amsart}
 
\title[Charged Cosmological Dust Solutions]{Charged Cosmological Dust
Solutions of the Coupled Einstein and Maxwell Equations
}
\author{Joel Spruck}\thanks{Research supported in part
by NSF grant DMS0904009}
\address{Department of Mathematics, Johns Hopkins University, Baltimore, MD 21218}
\author{Yisong Yang}
\address{Department of Mathematics,  Polytechnic Institute of New York University, Brooklyn, NY 11201}

\date{}
\newcommand{\bfR}{{\Bbb R}}

\newcommand{\goto}{\rightarrow}

\newtheorem{theorem}{Theorem}[section]
\newtheorem{lemma}[theorem]{Lemma}
\newtheorem{proposition}[theorem]{Proposition}
\newtheorem{corollary}[theorem]{Corollary}

\theoremstyle{definition}

\theoremstyle{remark}
\newtheorem{remark}[theorem]{Remark}

\newcommand{\dd}{\mbox{d}}\newcommand{\vph}{\varphi}
\newcommand{\ee}{\end{equation}}
\newcommand{\be}{\begin{equation}}\newcommand{\bea}{\begin{eqnarray}}
\newcommand{\eea}{\end{eqnarray}}

\newcommand{\pa}{\partial}\newcommand{\Om}{\Omega}

\newcommand{\nn}{\nonumber}

\newcommand{\om}{\omega}

\newcommand{\lm}{\lambda}

\begin{document}
\maketitle
\begin{center}
{ \small \em Dedicated to Louis Nirenberg
on the occasion of his 85th birthday}\\
\end{center}

\medskip 

\begin{abstract}
It is well known through the work of Majumdar, Papapetrou, Hartle, and Hawking that the coupled
Einstein and Maxwell equations admit a static multiple blackhole solution representing a balanced equilibrium state of finitely many point charges.  This is  a result of the exact cancellation of gravitational attraction and electric repulsion under an explicit condition on the mass and charge ratio.
The resulting system of particles,  known as an {\em extremely charged} dust, gives rise to examples of spacetimes with naked singularities. In this paper, we consider the continuous limit of the Majumdar--Papapetrou--Hartle--Hawking solution modeling a space occupied by an extended distribution of extremely charged dust. We show that for a given smooth distribution of matter of finite ADM mass there is a continuous family of smooth solutions realizing asymptotically flat space metrics.
\end{abstract}
\medskip

\section{Introduction}
The purpose of this paper is to establish the existence of an infinite family of smooth solutions to the 
statically coupled Einstein and Maxwell equations within the Majumdar \cite{Maj} and Papapetrou \cite{Pap}
metric describing the gravitational and electromagnetic fields generated from an extremely charged distribution
of cosmological dust. 

Consider a system of massive particles carrying electric charges of the same sign. In Newtonian theory, the 
inter-particle gravitational attraction and Coulomb repulsion both follow inverse square-power laws so that 
a perfect cancellation of the forces is reached when the ratios of masses and charges of the particles satisfy a balancing
condition. Under such a condition, a system of particles, as a distribution of charged dust, is referred to as {\em extremely charged}. Thus, due to the balanced forces, a static star of any geometric shape may be formed with an extremely charged dust. It is obviously
important to know whether the same conclusion may be reached in the context of general relativity when the Einstein equations are coupled with the Maxwell
equations. The earliest study of the Einstein--Maxwell equations was carried out by Reissner and Nordstr\"{o}m 
\cite{Wald} who obtained a static solution to the coupled field equations in empty space, which corresponds to the gravitational field of a charged, 
non-rotating, spherically symmetric massive body and generalizes the Schwarzschild blackhole metric. Since the Reissner--Nordstr\"{o}m solution concerns only
an isolated single particle, it gives no clue to the above extremely charged dust question. At roughly the same time as Reissner and Nordstr\"{o}m,  Weyl \cite{Weyl} classified 
all Einstein metrics induced by any static axially symmetric distribution of matter and charge. The work of Weyl work did not generate as much interest as
those by Schwarzschild, Reissner, and Nordstr\"{o}m, due to its geometric restriction of axial symmetry. A dramatic development came in 1947 when Majumdar \cite{Maj} and
Papapetrou \cite{Pap} independently extended Weyl's work to include an arbitrarily distributed system of extremely charged particles. In particular, the Majumdar--Papapetrou
formulation gives rise to the electrostatic solutions of the Einstein--Maxwell equations without any symmetry restriction, which marvelously generalizes the Newtonian theory
to the relativistic situation of multiple charged blackholes \cite{HH}. It has also been argued that, although astrophysical bodies are electrically neutral within a reasonable approximation,
the Majumdar--Papapetrou solutions may provide simple quasi-static analogies for complex dynamical processes preventing asymmetries
in gravitational collapses or collision of blackholes \cite{IW}.
More recently, it has been recognized that the Majumdar--Papapetrou solution allows a natural extension to all higher dimensions \cite{LZ,LZ2} which is known to be 
important to issues concerning unification of gravity with other forces in nature \cite{Myers,MPerry,Peet,RS1,RS2}.

Within the Majumdar--Papapetrou framework of an extremely charged discrete dust model, 
the coupled Einstein--Maxwell equations reduce to a Poisson equation in the vacuum space so that its solution represents a multiply distributed point charges with masses \cite{HH}. When a continuous distribution of extremely charged matter is considered, the field equations reduce to a semilinear elliptic equation.

However, in the literature for the continuous case, much attention has been given to dust models with spherical star or shell structures 
\cite{Cho,Gurses,GursesH,H,H2,H3,Ivanov,K,LZ2,V} with the focus on explicit construction of solutions.
There has not yet been any study of  the existence problem of the governing elliptic equation when the matter distribution is arbitrarily prescribed nor of
 the transition process from the continuous model to the discrete model. The purpose of the present paper is to establish an existence theory for this important gravitational system.

In the next section, we follow \cite{LZ} to lay out the governing elliptic equation for extremely charged dusts with a continuous matter distribution,
within the framework of Majumdar--Papapetrou. In Section 3, we consider the condition to
be imposed on the mass or charge density which ensures a finite ADM mass. In Section 4, we use the methods in Ni \cite{Ni} to establish an existence theory for the equation. 
We show that, in fact, the Majumdar--Papapetrou equation allows a continuous monotone family of finite ADM mass solutions labeled by their asymptotic values. In Section 5, we
derive some asymptotic estimates for the solutions obtained which lead to the expected asymptotic flatness of the gravitational metrics. In Section 6, we develop an existence theory
using an energy method \cite{ES,Ding} that requires much weaker conditions on the decay rate of the mass or charge density function. In Section 7, we present some nonexistence results. In particular, we show that when the asymptotic value of a conformal metric is larger than some $\beta_0>0$,  there can be no solution (whereas there are solutions when the asymptotic value is less than $\beta_0$). In the special situation when the mass or charge density function is radially symmetric, the set of permissible positive asymptotic values is precisely of the form $(0,\beta_0]$. This result complements the conditions in our existence theorems presented in Sections 4 and 6. 
Finally in Section 8, we draw our conclusions.

\section{The Majumdar--Papapetrou metric}
\setcounter{equation}{0}

Consider the $d=n+1$ dimensional Minkowski spacetime of signature $(-+\cdots+)$. Following \cite{LZ}, the metric element is given by
\be \label{2.1}
\dd s^2=g_{\mu\nu}\dd x^\mu\dd x^\nu=-V\dd t^2+h_{ij}\dd x^i\dd x^j,\quad \mu,\nu=0,1,\cdots,n,\quad i,j=1,\cdots,n.
\ee
Here and in the sequel, since we are interested in static solutions, all field variables depend on
the space coordinates ($x^i$) only. The universal gravitational constant and the speed of light are both taken to be unity. Using $\nabla_\mu$  to denote the covariant
derivative, $A_\mu$ the electromagnetic gauge potential, and $F_{\mu\nu}=\nabla_\mu A_\nu-\nabla_\nu A_\mu$ the electromagnetic field, we
can write the electromagnetic energy-momentum tensor
as
\be
4\pi E_{\mu\nu}=F_\mu^\alpha F_{\nu\alpha} -\frac14 g_{\mu\nu} F_{\alpha\beta}F^{\alpha\beta}.
\ee
Moreover, let $U_\mu$ be the underlying relativistic matter fluid $d$-velocity and $\rho_e$ and $ \rho_m$ the associated electric charge and
mass densities, respectively. Then the current density is
expressed as 
\be 
J_\mu =\rho_e U_\mu,
\ee
and the material energy-momentum tensor is written as
\be 
T_{\mu\nu}=\rho_m U_\mu U_\nu +M_{\mu\nu},
\ee
where $M_{\mu\nu}$ is the perfect-fluid, pressure-$p$, stress tensor given by
\be 
M_{\mu\nu}=p(U_\mu U_\nu+g_{\mu\nu}).
\ee
With the above notation, along with the Einstein tensor $G_{\mu\nu}=R_{\mu\nu}-\frac12 g_{\mu\nu} R$ represented in terms of the Ricci tensor $R_{\mu\nu}$ and
Ricci scalar curvature $R$, the coupled Einstein and Maxwell equations are
\bea 
G_{\mu\nu}&=& 8\pi (T_{\mu\nu}+E_{\mu\nu}),\\
\nabla_\nu F^{\mu\nu}&=&4\pi J^\mu.
\eea
As usual, the Weyl--Majumdar--Papapetrou ansatz in higher dimensions is generalized to be 
\be 
A_\mu=\phi\delta^0_\mu,\quad U_\mu=-\sqrt{V}\delta^0_\mu.
\ee
Thus, the conservation law $\nabla_\nu (T^{\mu\nu}+E^{\mu\nu})=0$, which is the relativistic analogue to the Euler equation, reads \cite{G,LZ}
\be 
\pa_i p+\frac1{2V}(\rho_m+p)\pa_i V-\frac1{\sqrt{V}}\rho_e\pa_i\phi=0,\quad i=1,\cdots,n,
\ee
which takes, after substituting the Weyl-type relation $V=V(\phi)$, the compressed form
\be \label{2.9}
\pa_i p +\bigg((\rho_m+p)\frac{V'}{2V}-\frac{\rho_e}{\sqrt{V}}\bigg)\pa_i\phi=0,\quad i=1,\cdots,n.
\ee
On the other hand, the higher dimensional generalization of the Majumdar--Papapetrou condition obtained in \cite{LZ} reads
\be \label{2.10}
V(\phi)=\bigg(a\pm\sqrt{\frac{2(n-2)}{n-1}}\phi\bigg)^2,
\ee
for some positive constant $a$ and the extremely charged perfect fluid condition \cite{LZ}, which follows from the compatibility of the
Einstein and Maxwell equations, is
\be \label{2.11}
\rho_e=4\frac{\sqrt{V}}{V'}\bigg(\frac{n-2}{n-1}\rho_m+\frac n{n-1}p\bigg).
\ee
As a consequence of (\ref{2.10}) and (\ref{2.11}), one arrives \cite{LZ} at
\be\label{2.13}
\rho_e=\pm\sqrt{\frac{2(n-2)}{n-1}}\bigg(\rho_m+\frac{n}{n-2}p\bigg).
\ee
In view of (\ref{2.9}), (\ref{2.10}), and (\ref{2.13}), it is seen that there holds $(n-2)V \pa_i p=p \pa_i V$, which, upon an integration,  gives us the relation  \cite{LZ}
\be 
p=C V^{\frac1{n-2}},
\ee
where $C$ is a constant. In a perfect-fluid star, we have $p=0$ at the boundary surface of the star. Thus, we arrive at $C=0$. In other words, $p=0$ everywhere and
the fluid becomes a dust. In particular, we are led to the following general dimensional extremely charged dust condition \cite{LZ} extending that of the Majumdar--Papapetrou
solution given as
\be \label{em}
\rho_e=\pm\sqrt{\frac{2(n-2)}{n-1}}\rho_m.
\ee
In the bottom situation with $n=3$, we have the classical condition for the extremely charged cosmological dust
\be 
|\rho_e|=\rho_m.
\ee

Choose $U>0$ and set
\be 
V=\frac1{U^2}.
\ee
Following \cite{Myers} and \cite{LZ}, we rewrite the metric element (\ref{2.1}) diagonally as
\be \label{2.17}
\dd s^2=-U^{-2}\dd t^2+U^{\frac2{n-2}}\delta_{ij}\dd x^i\dd x^j.
\ee
Then, under the condition (\ref{em}), the Einstein--Maxwell equations are reduced to the following single nonlinear elliptic equation \cite{LZ}
\be \label{MP}
\Delta U+8\pi\bigg(\frac{n-2}{n-1}\bigg)\rho_m(x) U^{\frac n{n-2}}=0,\quad x\in\bfR^n.
\ee

A solution of this equation will give rise to gravitational and electric forces generated from the extremely charged cosmological dust distribution. 
In the subsequent sections, we aim at
constructing solutions of the equation.

\section{An ADM mass calculation}
\setcounter{equation}{0}

Recall that the ADM mass \cite{ADM,Sch} of the space $(\bfR^n,\{h_{ij}\})$ housed within the Minkowski spacetime of the
 metric element (\ref{2.1}) is given by the formula
\be \label{3.1}
m_{\mbox{ADM}}=\frac1{2(n-1)\om_{n-1}}\lim_{r\to\infty}\int_{\pa B_r}(\pa_i h_{ij}-\pa_j h_{ii})\nu^j\,\dd S_r,
\ee
where $\dd S_r$ is the area element of $\pa B_r$ ($B_r=\{x\in\bfR^n\,|\,|x|<r\}$), $\nu$ denotes the outnormal vector to $\pa B_r$,  $\om_k$ is
the surface area of the standard unit sphere $S^k$, and the metric $\{h_{ij}\}$ is asymptotically Euclidean
satisfying
\be \label{3.2}
h_{ij}=c\delta_{ij}+\mbox{O}(|x|^{-(n-2)}),\quad \pa_k g_{ij}=\mbox{O}(|x|^{-(n-1)}),\quad \pa_\ell\pa_k g_{ij}=\mbox{O}(|x|^{-n}),
\ee
for $|x|\to\infty$ with $c>0$. Inserting (\ref{2.17}) into (\ref{3.1}), integrating, and using (\ref{MP}), we
obtain
\bea\label{ADM}
m_{\mbox{ADM}}&=&-\frac1{2\om_{n-1}}\lim_{r\to\infty}\int_{\pa B_r}\bigg(\frac{\pa}{\pa \nu}U^{\frac2{n-2}}\bigg)\,\dd S_r\nn\\
&=&-\frac1{2\om_{n-1}}\int_{\bfR^n}(\Delta U^{\frac2{n-2}})\,\dd x\nn\\
&=&\frac1{\om_{n-1}}\int_{\bfR^n}\bigg(\frac{8\pi\rho_m}{(n-1)} U^{\frac4{n-2}}+\frac{2(n-4)}{(n-2)^2} U^{\frac{2(3-n)}{n-2}}|\nabla U|^2\bigg)\,\dd x.
\eea
From (\ref{ADM}), we clearly
see that, when $n\geq4$, $m_{\mbox{ADM}}\geq0$ and $m_{\mbox{ADM}}=0$ if and only if $\rho_m\equiv0$ and the conformal metric is flat, $U=$constant. This statement may be viewed
as a much simplified (Majumdar--Papapetrou) version of the positive mass theorem \cite{PT,SY1,SY2,SY3,SY4,W}. Curiously, though, (\ref{ADM}) is not as clear in the classical dimension
$n=3$. Thus, we need to take a different path to recognize $m_{\mbox{ADM}}$.

Since (\ref{3.2}) is translated into 
\be \label{3.4}
U=U_\infty+\mbox{O}(|x|^{-(n-2)}),\quad \pa_i U =\mbox{O}(|x|^{-(n-1)}),\quad \pa_i\pa_j U=\mbox{O}(|x|^{-n}),
\ee
 as $|x|\to\infty$ for some $U_\infty>0$, we have, in view of (\ref{MP}) again, and the decay estimates stated in (\ref{3.4}),
\bea\label{ADM2}
m_{\mbox{ADM}}
&=&
 -\frac1{(n-2)\omega_{n-1}} \lim_{r\to\infty} \int_{\partial B_r}  U^{\frac{4-n}{n-2}}\frac{\pa U}{\pa\nu} \,\dd S_r\nn\\
&=& -\frac1{(n-2)\omega_{n-1} }\lim_{r\to\infty} \left(\int_{\partial B_r}\left[ U^{\frac{4-n}{n-2} }-U_\infty^{\frac{4-n}{n-2} }\right]\frac{\pa U}{\pa\nu} \,\dd S_r 
     +  \int_{\partial B_r}   U_\infty^{\frac{4-n}{n-2} }\frac{\pa U}{\pa\nu}\,\dd S_r\right) \nn\\
&=&     \frac{8\pi}{(n-1)\omega_{n-1} }    U_\infty^{\frac{4-n}{n-2} } \int_{\bfR^n} \rho_m(x) U^{\frac{n}{n-2}}(x)\,\dd x,
\eea
which is indeed positive-definite and vanishes only when $\rho_m\equiv0$ and the metric becomes globally flat. Thus, we have arrived at a rather transparent (simplified) version of the positive
mass theorem within the Majumdar--Papapetrou metric in all dimensions, $n\geq3$.

In view of (\ref{3.4}) and (\ref{ADM2}), we obtain the finite integral requirement
\be \label{3.5}
\int_{\bfR^n}\rho_m(x)\,\dd x<\infty.
\ee

In the subsequent sections, we will construct solutions of  the Majumdar--Papapetrou equation (\ref{MP}), observing the condition (\ref{3.5}).

\section{Existence of solutions by a sub-supersolution method}
\setcounter{equation}{0}

Following the above discussion, we rewrite the Majumdar--Papapetrou equation over $\bfR^n$ ($n\geq 3$) in the compressed form
\be\label{1}
\Delta u + K(x) u^{\frac n{n-2}}=0,\quad x\in\bfR^n,
\ee
where $K(x)$ is a smooth nonnegative function in  $\bfR^n$ which is not identically zero and lies in $L^1(\bfR^n)$.
That is,
\be
\label{KL1} 
\int_{\bfR^n} K(x)\,\dd x<\infty.
\ee
So it is natural to assume that
\be \label{3}
K(x)=\mbox{O}(|x|^{-\ell})\quad \mbox{as }|x|\to\infty
\ee
where $\ell$ is large enough. For existence, a convenient condition is
\be \label{4}
\ell>2.
\ee
From (\ref{3}) and (\ref{4}), we may assume that there is a function $K_0(r)>0$ ($r=|x|$) such that
\be \label{5}
K(x)\leq K_0(r),\quad r>0;\quad K_0(r)=\mbox{O}(r^{-\ell})\quad\mbox{as } r\to\infty;\quad\ell >2.
\ee

We may adapt the method of sub- and supersolutions developed in \cite{Ni}. For this purpose, consider
\bea\label{6}
u_{rr}+\frac{(n-1)} r u_r &=& - K_0(r) u^{\frac n{n-2}},\quad r>0,\\
u(0)&=&\alpha>0,\quad u_r(0)=0.\label{7}
\eea

It is well known that (\ref{6})--(\ref{7}) can be solved locally over $[0,R)$ where $R>0$ is finite or infinite. Integrating
(\ref{6})--(\ref{7}), we get
\be 
r^{n-1} u_r(r) =-\int_0^r \rho^{n-1} K_0(\rho) u^{\frac n{n-2}}(\rho)\,\dd\rho,\quad 0<r<R.
\ee
Hence $u(r)$ decreases in $[0,R)$ if $u(r)$ remains positive-valued.

Following \cite{Ni}, we consider a test function $v$ defined by
\be 
v(r)=\frac1{(1+r^2)^a},\quad a>0.
\ee
Then $v(0)=1$, $v_r(0)=0$, and
\be 
v_{rr}+\frac{(n-1)} r v_r = -\bigg(n-\frac{2(a+1)r^2}{(1+r^2)}\bigg)\frac{2a}{(1+r^2)^{a+1}},\quad r>0.
\ee
Thus, when
\be \label{11}
n-2(a+1)\geq 0\quad\mbox{or}\quad a\leq\frac{n-2}2,
\ee
$v$ is superharmonic, $\Delta v<0$. We shall maintain the condition (\ref{11}) in the sequel.

Let $u$ be the unique local solution of (\ref{6})--(\ref{7}) which remains positive over its maximal interval $[0,R)$. That is, the solution exists 
 and stays positive in $[0,R)$ for some $R>0$ so that $u(R)=0$ if $R$ is finite.

Define $\varphi(r)=u(r)-\alpha v(r)$, $r\in[0,R)$. Then $\varphi(0)=0,\varphi_r(0)=0$, and $\varphi$ satisfies
\bea 
\vph_{rr}+\frac{(n-1)}r\vph_r&=&-K_0(r) u^{\frac n{n-2}} +\bigg(n-\frac{2(a+1)r^2}{(1+r^2)}\bigg)\frac{2a\alpha}{(1+r^2)^{a+1}}\nn\\
&\geq&-K_0(r)\alpha^{\frac n{n-2}} +\bigg(n-\frac{2(a+1)r^2}{(1+r^2)}\bigg)\frac{2a\alpha}{(1+r^2)^{a+1}}\nn\\
&=&\alpha\bigg[\bigg(n-\frac{2(a+1)r^2}{(1+r^2)}\bigg)\frac{2a}{(1+r^2)^{a+1}}-K_0(r)\alpha^{\frac 2{n-2}}\bigg]\nn\\
&\equiv& Q(r),\quad r<R.\label{12}
\eea

We want to make $Q(r)>0$ for all $r>0$. To do so, we set 
\be \label{13}
2(a+1)\leq\ell
\ee
 for $\ell>2$ given in (\ref{5}). Combining  (\ref{11}) and (\ref{13}), we may assume
\be \label{14}
a\leq\frac12\min\{n-2,\ell-2\}.
\ee

In (\ref{5}), we may assume without loss of generality that there is a constant $C_\ell>0$ such that
\be \label{15}
K_0(r)\leq\frac{C_\ell}{(1+r)^\ell},\quad r\geq0.
\ee 

In view of (\ref{11})--(\ref{15}), we can find some $\alpha_0>0$ such that
\be \label{16}
Q(r)>0,\quad r>0,\quad \alpha\in (0,\alpha_0).
\ee

From (\ref{12}) and (\ref{16}), we have as before,
\be\label{17}
r^{n-1} \vph_r(r) >\int_0^r \rho^{n-1} Q(\rho)\,\dd\rho>0,\quad 0<r<R.
\ee
In particular, $\vph(r)$ increases for $0<r<R$ and $\vph(0)=0$ implies 
\be 
\lim_{r\to R}\vph(r)>0\quad\mbox{or}\quad \lim_{r\to R} u(r)=\lim_{r\to R}\vph(r)+\alpha  \lim_{r\to R}v(r)>0.
\ee
This result establishes $R=\infty$. In other words, there is some $\alpha_0>0$ such that for each $\alpha\in (0,\alpha_0)$ the problem (\ref{6})--(\ref{7}) has a unique solution, say $u_\alpha(r)$ over $0<r<\infty$ so that
$\vph_\alpha (r)=u_\alpha(r)-\alpha v(r)$ increases and $u_\alpha(r)$ stays positive but decreases for all $r>0$. Consequently, we have the positive finite limit
\be \label{19}
\lim_{r\to\infty} u_\alpha (r)=\lim_{r\to\infty}\vph(r)\equiv\beta_\alpha>0.
\ee

As a solution of $\Delta u +K_0 u^{\frac n{n-2}}=0$ in $\bfR^n$, $u_\alpha$ is a supersolution of (\ref{1}) because
the property $u_\alpha>0$ and (\ref{5}) imply that
\be 
-\Delta u_\alpha = K_0 (|x|) u^{\frac n{n-2}}_\alpha\geq K(x) u^{\frac n{n-2}}_\alpha,\quad x\in\bfR^n.
\ee
Set $v_\alpha(x)=\beta_\alpha$. We see that $v_\alpha$ is a subsolution of (\ref{1}) since $K(x)\geq0$. However, we have seen that $v_\alpha<u_\alpha$.
As a consequence, we have
\begin{lemma} \label{sub-sup}
Equation (\ref{1}) has a solution $u$ satisfying
\bea \label{21}
\beta_\alpha=v_\alpha\leq u\leq u_\alpha\leq\alpha,\quad x\in\bfR^n\\
u(x)\to\beta_\alpha\quad\mbox{as }|x|\to\infty.
\eea
Moreover, $u$ is the unique minimal positive solution of (\ref{1}) in the sense that if $w$ is any other solution  satisfying $w(x)\to\beta_\alpha\,\,\mbox{as }|x|\to\infty$,
then $w \geq u$.
\end{lemma}
\begin{proof} We briefly sketch the proof since it is standard. We first  solve
\begin{eqnarray*}
-\Delta u_{j+1}&=&K u_j^{\frac{n}{n-2}} \,\, \mbox{in $B_R(0)$}\\
u_{j+1}&=&\beta_{\alpha}\,\,\mbox{on $\partial B_R(0)$}\\
u_0 &=&v_{\alpha}
\end{eqnarray*}
and observe that 
\[ v_{\alpha} \leq u_j \leq v_{j+1} <u_{\alpha}\,\,\mbox{(respectively $w$)}.\]
(Here we used that $w \geq \beta_{\alpha}$; see section 7 for a simple proof.)
Hence $u_{R}=\lim_{j \to \infty}u_j$ solves
\begin{eqnarray*}
-\Delta u_R&=&K u_R^{\frac{n}{n-2}} \,\, \mbox{in $B_R(0)$}\\
u_R&=&\beta_{\alpha}\,\,\mbox{on $\partial B_R(0)$}
\end{eqnarray*}
Finally, in the limit  as $R$ tends to infinity, we obtain the unique minimal solution $u$, of (\ref{1}) satisfying
 \[v_{\alpha} \leq u \leq u_{\alpha} \,\, \mbox{(respectively $w$)}.\]
\end{proof}

As a by-product of (\ref{19}) and (\ref{21}), we have the immediate result
\be 
u(x)\to\beta_\alpha\quad\mbox{as }|x|\to\infty.
\ee

Next observe that 
\be \label{s1}
u_\alpha^{(\beta)} =\left(\frac\beta{\beta_\alpha}\right) u_\alpha,\quad 0<\beta\leq\beta_\alpha,
\ee
 satisfies 
\be \label{s2}
u_\alpha^{(\beta)} \geq \beta,\quad u_\alpha^{(\beta)}(x) \to\beta\mbox{ as }|x|\to\infty,
\ee
and
\be \label{s3}
-\Delta u_\alpha^{(\beta)}  =\left(\frac{\beta}{\beta_\alpha}\right)^{-\frac2{n-2}}K_0 (u_\alpha^{(\beta)})^{\frac{n}{n-2}} \geq K (u_\alpha^{(\beta)})^{\frac{n}{n-2}}.
\ee
Combining (\ref{s1})--(\ref{s3}), we see that we have obtained a sub- and supersolution pair, consisting of $v^{(\beta)}_\alpha\equiv\beta$ and $u_\alpha^{(\beta)}$,
satisfying $v^{(\beta)}_\alpha\leq u^{(\beta)}_\alpha$ for each $0<\beta\leq\beta_\alpha$. In particular, using the above procedure, we see that there is a minimal
positive solution $u^{(\beta)}(x)$ for any $0<\beta \leq \beta_{\alpha}$ which depends continuously on
$\beta$.

\medskip

In summary, we have

\begin{theorem}\label{theorem1} Suppose that the function $K(x)\geq0$ satisfies 
\be 
K(x)=\mbox{O}(|x|^{-\ell}),\quad \mbox{as }|x|\to\infty,\quad \ell>2. 
\ee
Then there is a positive number $\beta_0>0$ such that the Majumdar--Papapetrou equation (\ref{1}) has a continuum of minimal positive solutions
$\{u^{(\beta)}\,|\,0<\beta<\beta_0\}$ satisfying
\be\label{}
\lim_{|x|\to\infty} u^{(\beta)}(x)=\beta,\quad u^{(\beta)}\geq\beta.
\ee
 Moreover $u^{(\beta)}$ depends continuously in $\beta$ and is
monotone increasing in $\beta$.
\end{theorem}
In Section 7 we will show that $\beta_0<\infty$.

We next consider the asymptotic decay estimates of the solutions obtained in Theorem \ref{theorem1} as $|x|\to\infty$.

\section{Asymptotic estimates}
\setcounter{equation}{0}

We need the following somewhat standard result concerning a Newton-type potential integral.

\begin{lemma} \label{lemma2}
Define the function $v(x)$ by the Newton-type integral
\be\label{newton}
v(x)=\int_{\bfR^n}\frac{f(y)}{|y-x|^{\sigma}}\,\dd y,
\ee
where $f(x)=\mbox{\rm O}(|x|^{-\gamma})$ as $|x|\to\infty$ for some constants $\gamma$ and $\sigma<n$ and
\be 
\int_{\bfR^n}|f(x)|\,\dd x<\infty. 
\ee 
Then we have
\be \label{v}
v(x)=\mbox{\rm O}(|x|^{-\delta})\quad\mbox{as }|x|\to\infty,
\ee
where $\delta=\min\{\sigma,\gamma+\sigma-n\}$, provided that the exponents $\gamma$ and $\sigma$ satisfy the condition 
\be \label{gamma}
\gamma+\sigma>n.
\ee
\end{lemma}
\begin{proof}
For fixed $x\in\bfR^n$, $x\neq 0$, we decompose $\bfR^n$ into $\bfR^n=\Om_1\cup\Om_2\cup\Om_3$ where
\bea 
\Om_1&=&\bigg\{ y\in\bfR^n\,\bigg|\, |y-x|\leq \frac{|x|}2\bigg\},\nn\\
\Om_2&=&\bigg\{ y\in\bfR^n\,\bigg|\, \frac{|x|}2\leq |y-x|\leq 2{|x|}\bigg\},\nn\\
\Om_3&=&\bigg\{ y\in\bfR^n\,\bigg|\, |y-x|\geq2{|x|}\bigg\},\nn
\eea
which results in $v(x)=v_1(x)+v_2(x)+v_3(x)$ with
\be 
v_i(x)=\int_{\Om_i}\frac{f(y)}{|y-x|^{\sigma}}\,\dd y,\quad i=1,2,3.
\ee
On the other hand, using the condition on $f(x)$, we have the bound 
\be\label{}
|f(y)|\leq\frac{C}{1+|y|^\gamma}\leq
\frac{C}{1+(|x|/2)^\gamma}\leq\frac{2^\gamma C}{1+|x|^\gamma},\quad y\in\Om_1,
\ee
in view of $|x|-|y|\leq |x|/2$ for $y\in\Om_1$. Therefore,
\bea \label{v1}
|v_1(x)|&\leq& \frac{2^\gamma C}{(1+|x|^\gamma)}\int_{|y-x|\leq\frac{|x|}2}\frac{\dd y}{|y-x|^{\sigma}}\nn\\
&=&\mbox{O}(|x|^{-(\gamma+\sigma-n)})\quad \mbox{as }|x|\to\infty.
\eea
For $y\in\Om_2$, we have
\bea\label{v2}
|v_2(x)|&\leq&\frac{2^\sigma}{|x|^\sigma}\int_{\frac12|x|\leq|y-x|\leq2|x|}|f(y)|\,\dd y\nn\\
&\leq&\frac{2^\sigma}{|x|^\sigma}\int_{\bfR^n}|f(y)|\,\dd y\nn\\
&=&\mbox{O}(|x|^{-\sigma})\quad \mbox{as }|x|\to\infty.
\eea
 Furthermore, using $| |x-y|-|x||\leq |y|$, we have
\bea\label{v3}
|v_3(x)|&\leq& C\int_{|y-x|\geq2|x|}\frac{\dd y}{|y-x|^\sigma(1+|y|^\gamma)}\nn\\
&\leq&{C}\int_{|y-x|\geq 2|x|}\frac{\dd y}{|y-x|^\sigma(1+[|y-x|-|x|]^\gamma)}\nn\\
&\leq& C\int_{|y-x|\geq 2|x|}\frac{\dd y}{|y-x|^\sigma (1+|y-x|^\gamma/2^\gamma)}\nn\\
&=&\mbox{O}(|x|^{-(\gamma+\sigma-n)})\quad\mbox{as }|x|\to\infty,
\eea
using $||y-x|-|x||\geq|y-x|/2$, which is the same as stated in (\ref{v1}) provided that $\gamma+\sigma>n$ holds. 

Hence, summarizing (\ref{v1})--(\ref{v3}), we see that (\ref{v}) is established.
\end{proof}

We now turn our attention to (\ref{1}). Let $u$ be a positive solution of (\ref{1}) produced in Theorem \ref{theorem1}. Then there is a constant $\beta>0$
such that
\be \label{ubc}
\lim_{|x|\to\infty} u(x)=\beta.
\ee

Let $w(x)$ be given by the Newton potential associated with (\ref{1}), i.e.,
\be\label{w}
w(x)=\frac1{n(n-2)\Om_n}\int_{\bfR^n}\frac{K(y) u^{\frac{n}{n-2}}(y)}{|x-y|^{n-2}}\,\dd y,
\ee
where $\Om_n=2\pi^{n/2}/n\Gamma(n/2)$ is the volume of the unit ball in $\bfR^n$. Applying Lemma \ref{lemma2} with
$\sigma=n-2$ to $w(x)$ and noting (\ref{ubc}), we see that, with $\ell=\gamma$,
the conditions (\ref{4}) and (\ref{gamma}) coincide. Therefore, we obtain
\be\label{wbc}
w(x)=\mbox{O}(|x|^{-m})\quad\mbox{as }|x|\to\infty,\quad m=\min\{n-2,\ell-2\}.
\ee

On the other hand, since $h=u-w$ is a bounded harmonic function in view of (\ref{ubc}) and (\ref{wbc}), it must be a constant, which must be given by (\ref{ubc}). In other words, we have
established the relation
\be \label{ubeta}
u(x)=\beta+\frac1{n(n-2)\Om_n}\int_{\bfR^n}\frac{K(y) u^{\frac n{n-2}}(y)}{|x-y|^{n-2}}\,\dd y,\quad x\in\bfR^n.
\ee
 Differentiating (\ref{ubeta}), we have
\bea\label{Dubc}
|\pa_i u(x)|&\leq&\frac1{n\Om_n}\int_{\bfR^n}\frac{K(y)u^{\frac n{n-2}}(y)}{|x-y|^{n-1}}\,\dd y\nn\\
&=&\mbox{O}(|x|^{-( m+1)})\quad\mbox{as }|x|\to\infty,\quad i=1,\cdots,n,
\eea
in view of Lemma \ref{lemma2} again.

In order to get the decay estimate for $D^2 u$ near infinity, we may be tempted to differentiate (\ref{ubeta}) twice and apply Lemma \ref{lemma2} again. However, such an 
approach will not work because of the technical restriction $\sigma<n$ for the Newton-type integral (\ref{newton}) stated in Lemma \ref{lemma2}.
To overcome this difficulty, we set $|x|=R$, assume that $R$ is sufficiently large, and rewrite (\ref{ubeta}) as
\be\label{uUV}
u(x)=\beta +U(x)+V(x),
\ee
where
\be
U(x)=\int_{|y|\leq 2R} \Gamma(x-y)f(y)\,\dd y,\quad V(x)=\int_{|y|\geq 2R}\Gamma(x-y)f(y)\,\dd y,\label{UV}
\ee
with
$\Gamma(z)=1/n(n-2)\Om_n |z|^{2-n}$ and $f(y)=-K(y) u^{\frac n{n-2}}(y)$. Then $|\pa_i\pa_j \Gamma(z)|\leq1/\Om_n |z|^n$.

First, we have by using $|x-y|\geq |y|-R$ and $f(y)=\mbox{O}(|y|^{-\ell})$ for $|y|\to\infty$,
\bea\label{ddV}
|\pa_i\pa_j V(x)|&\leq&\int_{|y|\geq 2R}|\pa_i\pa_j \Gamma(x-y) f(y)|\,\dd y\nn\\
&\leq& C\int_{|y|\geq 2R}\frac{|f(y)|}{|x-y|^n}\,\dd y\nn\\
&\leq& 2^n C\int_{|y|\geq 2R} |y|^{-(\ell +n)}\,\dd y \nn\\
&=& C_1 R^{-\ell},\quad i,j=1,\cdots,n.
\eea

In order to estimate the second derivatives of $U(x)$ in (\ref{uUV}), we extend $f(y)$ such that $f(y)=0$ for $|y|>2R$ and use Lemma 4.2 in \cite{GT} to represent $\pa_i\pa_j U(x)$ as
\bea\label{5.18}
\pa_i\pa_j U(x)
&=&\int_{|y|\leq 3R} \pa_i \pa_j \Gamma (x-y)(f(y)-f(x))\,\dd y
-f(x)\int_{|y|=3R}\pa_i\Gamma(x-y)\nu_j(y)\,\dd S_y\nn\\
&\equiv& U_1(x)+U_2(x),\quad i,j=1,\cdots,n,
\eea
where $(\nu_j(y))$ is the outnormal vector at $y$ over the sphere $\{|y|=3R\}$ and $\dd S_y$ is the area element.

In view of $|x-y|\geq||y|-|x||=2R$ for $|y|=3R$, we have
\bea \label{5.19}
|U_2(x)|&\leq&|f(x)|\int_{|y|=3R}|\pa_i\Gamma (x-y)\nu_j(y)|\,\dd S_y\nn\\
&\leq& C |f(x)|R^{-(n-1)}\int_{|y|=3R}\,\dd S_y\nn\\
&\leq&C_2 R^{-\ell}.
\eea

To estimate $U_1(x)$, we let $\Om_k$ ($k=1,2,3$) be defined earlier and write
\bea \label{5.20}
U_1(x)&=&\sum_{k=1}^3\int_{\Om_k\cap\{|y|\leq3R\}}\pa_i\pa_j\Gamma(x-y)(f(y)-f(x))\,\dd y \nn\\
&\equiv& W_1(x)+W_2(x)+W_3(x).
\eea
We impose the additional natural decay condition on $K$ as follows,
\be\label{DK} 
\pa_i K(x)=\mbox{O}(|x|^{-(\ell +1)})\quad\mbox{as }|x|\to\infty,\quad i=1,\cdots,n.
\ee
In view of (\ref{DK}), (\ref{Dubc}), and the relation $f=Ku^{\frac n{n-2}}$, we have
\be 
\pa_i f(x)=\mbox{O}(|x|^{-(\ell+1)})\quad\mbox{as }|x|\to\infty,\quad i=1,\cdots,n.
\ee
Since for $y\in\Om_1$, we have $|y|\leq|y-x|+|x|\leq 3R/2<2R$, we see that $W_1(x)$ satisfies the estimate
\bea \label{W1}
|W_1(x)|&\leq&\int_{\{|x-y|\leq R/2\}\cap\{|y|\leq 2R\}}|\pa_i\pa_j\Gamma(x-y)(f(x)-f(y))|\,\dd y\nn\\
&\leq& C\int_{|x-y|\leq R/2}\frac{|f(x)-f(y)|}{|x-y|^n}\,\dd y\nn\\
&\leq& C\int_{|x-y|\leq R/2}\frac{|\nabla f(\xi)|}{|x-y|^{n-1}}\,\dd y,
\eea
where $\xi=x+t(y-x)$ for some $t\in[0,1]$. Hence $|\xi|\geq|x|-|y-x|\geq R-R/2=R/2$ and $|\nabla f(\xi)|\leq C_3 R^{-(\ell+1)}$ for some constant $C_3>0$. Inserting this result into (\ref{W1})
and integrating,
we obtain
\be \label{5.24}
|W_1(x)|\leq C_4 R^{-\ell}.
\ee

To estimate the decay rate of $W_2(x)$, we first note that
\be \label{5.25}
\int_{\frac R2\leq|y-x|\leq 2R}|\pa_i\pa_j \Gamma(x-y)f(y)|\,\dd y\leq \frac{2^n}{\Om_n R^n}\int_{\bfR^n}|f(y)|\,\dd y.
\ee 
Moreover, in view of $f(x)=\mbox{O}(R^{-\ell})$,  we have
\bea \label{5.26}
\int_{\frac R2\leq|y-x|\leq 2R}|\pa_i\pa_j \Gamma(x-y)f(x)|\,\dd y&\leq&  \mbox{O}(R^{-\ell})\frac{2^n}{\Om_n R^n}\int_{\frac R2\leq|y-x|\leq 2R}\,\dd y\nn\\
&\leq& 2^{2n}\mbox{O}(R^{-\ell}).
\eea
Combining (\ref{5.25}) and (\ref{5.26}), we obtain
\be\label{5.27}
|W_2(x)|\leq C_5 R^{-\min\{\ell, n\}}.
\ee

For $W_3(x)$, we make a similar decomposition. First we have the estimate
\be \label{5.28}
\int_{|y-x|\geq 2R}|\pa_i\pa_j \Gamma(x-y)f(y)|\,\dd y\leq \frac1{2^n\Om_n R^n}\int_{\bfR^n}|f(y)|\,\dd y.
\ee
Furthermore, by virtue of $f(x)=\mbox{O}(R^{-\ell})$ again, we find
\bea \label{5.29}
\int_{\{|y-x|\geq 2R\}\cap\{|y|\leq 3R\}}|\pa_i\pa_j \Gamma(x-y)f(x)|\,\dd y&\leq& \frac{\mbox{O}(R^{-\ell})}{2^n\Om_n R^n}\int_{\{|y-x|\geq 2R\}\cap\{|y|\leq 3R\}}\,\dd y\nn\\
&\leq& \left(\frac32\right)^n \mbox{O}(R^{-\ell}).
\eea
Combining (\ref{5.28}) and (\ref{5.29}), we arrive at
\be\label{5.30}
|W_3(x)|\leq C_5 R^{-\min\{\ell,n\}}.
\ee

Consequently, we may now collect (\ref{5.19}), (\ref{5.24}), (\ref{5.27}), and (\ref{5.30}) and recall the decompositions (\ref{5.18}) and (\ref{5.20}) to obtain
the decay estimate
\be \label{5.31}
|\pa_i\pa_j U(x)|\leq C_6 |x|^{-\min\{\ell,n\}}\quad\mbox{as }|x|\to\infty,\quad i,j=1,\cdots,n.
\ee

Finally, inserting (\ref{5.31}) and (\ref{ddV}) into (\ref{uUV}), we see that the second derivatives of the solution $u(x)$ satisfy the same decay estimates as $U(x)$.

Summarizing the above discussion, we can state

\begin{theorem}
Let $u$ be a solution obtained in Theorem \ref{theorem1} with $u(x)\to\beta>0$ as $|x|\to\infty$ when the charge density function $K$ satisfies
the conditions (\ref{3}) and (\ref{4}). Then $u$ enjoys the decay estimates
\be\label{uDu}
u(x)-\beta =\mbox{O}(|x|^{-m}),\quad Du(x)=\mbox{O}(|x|^{-(m+1)}),\quad |x|\to\infty,
\ee
where $m=\min\{n-2,\ell-2\}$ and $D$ denotes a generic derivative.
Moreover, if the function $K$ satisfies the additional condition (\ref{DK}), then $D^2 u$ obeys $(D^2 u)(x)=\mbox{O}(|x|^{-(m+2)})$ as $|x|\to\infty$.
\end{theorem}

The asymptotic flatness condition (\ref{3.4}) may be ensured with the assumption $\ell =n$ such that $m=n-2$.

\section{Existence of solutions by an energy method}
\setcounter{equation}{0}

In this section, we shall develop an existence theory for the Majumdar--Papapetrou equation (\ref{1}) using an
energy method.

We assume that $K(x)$ satisfies (\ref{3}) with $\ell>1$.
If the constant $\beta \geq 0$ defined below is positive, we assume in addition that (\ref{KL1}) is fulfilled.

 Choose a sequence $\{\theta_k\}$ of positive constants decreasing to zero and define the Hilbert space
 \[ \mathcal{H}^1_k= \{v \in H^1(\bfR^n): \int_{\bfR^n}( |\nabla v|^2 + \theta_k  v^2)\,\dd x< \infty\},
\]
with inner product
\[ (v,w)_{k}=\int_{\bfR^n}( \nabla v \cdot \nabla w + \theta_k  vw)\,\dd x ~.
\]
and norm $\|v\|_k=(v,v)_k^{\frac12}$.  For later use, define also the Hilbert space $\mathcal{H}_K$
as the completion of the smooth functions of compact support under the inner product
\[( v,w)_{K}=\int_{\bfR^n}( \nabla v \cdot \nabla w + K vw)~\dd x ~.\]

We will be interested in the subset
\[ \mathcal{A}_k= \{v \in \mathcal{H}^1_k: v\geq 0,\,  \|v\|_k=1\}~.\]
Fix   $v_0 \in  C^{\infty}(\bfR^n) \cap \mathcal{H}^1_0 \cap L^{\infty}(\bfR^n),\,|| v_0||_0 \leq 1,\,  v_0 \geq 0$.  We define a sequence $v_k\in  \mathcal{A}_k , \, k\geq 1$ inductively  as follows:  for a fixed constant $\beta \geq 0$, let
$w_{k+1} \in \mathcal{H}^1_{k+1}$ be the unique solution of 
\be \label{joeleq1}
-\Delta w +\theta_{k+1}w = K (v_k +\beta)^{\frac{n}{n-2}} 
\ee
and set $v_{k+1}= \lambda_{k+1}w_{k+1}$ where $\lambda_{k+1}=\frac1{\|w_{k+1}\|_{k+1}}$. Then
\be \label{joeleq2}
 v_{k+1} \in   \mathcal{A}_{k+1}\,\,\mbox{and}\,\,   -\Delta v_{k+1}+ \theta_{k+1} v_{k+1} =
\lambda_{k+1}K (v_k+\beta)^{\frac{n}{n-2}}~. \ee

To see the legitimacy of the above procedure, we inductively assume that $v_k\in \mathcal{H}^1_k$ and $v_k\geq0$ some $k\geq1$. 

Recall the Gagliardo--Nirenberg inequality
\be\label{joelGN}
\left(\int_{\bfR^n} v^{\frac{2n}{n-2}} \,\dd x\right) ^ {\frac{n-2}n} \leq C \int_{\bfR^n} |\nabla v|^2 \,\dd x. 
\ee
As a consequence of (\ref{KL1}) and (\ref{joelGN}), we see that $K(v_k+\beta)^{\frac n{n-2}}\in L^2(\bfR^n)$ since
\bea 
 \int_{\bfR^n} K^2 (v_k+\beta)^{\frac{2n}{n-2}}\,\dd x &\leq& 
C_1\int_{\bfR^n} K^2(v_k^{\frac{2n}{n-2}}+\beta^{\frac{2n}{n-2}})\,\dd x\nn\\
&\leq&
 C_2\left(\int_{\bfR^n} |\nabla v_k|^2 \,\dd x\right)^{\frac{n}{n-2}} + C_3\int_{\bfR^n} K\,\dd x<\infty.\nn
\eea
Using the Lax--Milgram theorem, we see that (\ref{joeleq1}) has a unique solution, say $w_{k+1}$, in $\mathcal{H}^1_{k+1}$, as claimed. 

We need to show that $w_{k+1}\geq0$. To see this, we consider the weak form of (\ref{joeleq1}), that is,
\be \label{joel.weak}
\int_{\bfR^n} (\nabla w_{k+1} \cdot \nabla \phi + \theta_{k+1} w_{k+1} \phi)\,\dd x= \int_{\bfR^n} K(v_k+\beta)^{\frac n{n-2}} \phi \,\dd x,\quad\phi\in\mathcal{H}^1_{k+1}.
\ee
Let $m>0$ and observe that $\phi=(-m-w_{k+1})^{+} \in \mathcal{H}^1_{k+1}$. Inserting this choice of
$\phi$ into (\ref{joel.weak}) gives
\[ \int _{\{w_{k+1}<-m\}} \theta_{k+1} w_{k+1}\phi\, \dd x  \geq \int_{\bfR^n} K(v_k+\beta)^{\frac n{n-2}} \phi \,\dd x \geq 0,\]
which is impossible unless the set $\{w_{k+1}<-m\}$ is of measure zero. Since $m$ is arbitrary, we have shown that $w_{k+1}\geq0$ a.e., which establishes the legitimacy of the construction of
the sequence $\{v_k\}$ as stated.

We define 
\[\Phi(v)=\int_{\bfR^n} K(v^{+}+\beta)^{\frac{2n-2}{n-2}}\,\dd x~.\]
Note that for $v \in  \mathcal{A}_{k}$, we have the uniform bound
\bea\label{joelPhi}
\Phi(v) &\leq& 2^{\frac{n}{n-2}} \int_{\bfR^n}K(v^{\frac{2n-2}{n-2}}+\beta^{\frac{2n-2}{n-2}})\,\dd x\nn\\
 &\leq& 
C\left(\beta^{\frac{2n-2}{n-2}}\int_{\bfR^n}K\,\dd x+\|K\|_{L^n(\bfR^n)}\left(\int_{\bfR^n} |\nabla v|^2 \dd x\right)^{\frac{n-1}n}\right)\nn\\
&\leq& C\left(\beta^{\frac{2n-2}{n-2}}\int_{\bfR^n}K\,\dd x+\|K\|_{L^n(\bfR^n)}\right),
\eea
with a uniform constant $C>0$.

\begin{lemma}\label{joellem1} $\Phi(v_k) \leq \Phi(v_{k+1}),\, k=0,1, \ldots$\end{lemma}
\begin{proof} From (\ref{joeleq1}), we have
\bea \label{joeleq3}
&&\int_{\bfR^n}(\nabla v_{k+1}\cdot \nabla(v_{k+1}-v_k)+\theta_{k+1}v_{k+1}(v_{k+1}-v_k))\,\dd x\nn\\
&&= \lambda_{k+1} \left(\int_{\bfR^n}K(v_{k+1}+\beta)
(v_k+\beta)^{\frac{n}{n-2}}\,\dd x-\Phi(v_k)\right) \nn\\
&& \leq  \lambda_{k+1} \left(\frac{n-2}{2n-2} \Phi(v_{k+1})+\frac{n}{2n-2} \Phi(v_k)-\Phi(v_k)\right)\nn\\
&&=\frac{n-2}{2n-2} \lambda_{k+1}(\Phi(v_{k+1})-\Phi(v_k)).
\eea
However, the left-hand side of (\ref{joeleq3}) satisfies
\begin{eqnarray}
&&\int_{\bfR^n}(\nabla v_{k+1}\cdot \nabla(v_{k+1}-v_k)+\theta_{k+1}v_{k+1}(v_{k+1}-v_k))\,\dd x \nn\\
&&\geq \frac12( \|v_{k+1}\|_{k+1}^2 -\|v_k\|_k^2 +(\theta_k-\theta_{k+1})\|v_k\|_{L^2(\bfR^n)}^2) \geq 0.\label{joeleq1.8}
\end{eqnarray}
\end{proof}

\begin{corollary} \label{joelcor1}Assume that $\beta$ lies in the interval
 \be\label{joelbeta}
 0\leq \beta < \beta_0\equiv \left(\frac{\int_{\bfR^n} K v_0^{\frac{2n-2}{n-2}}\,\dd x}{ \int_{\bfR^n}K\,\dd x}\right)^{\frac{n-2}{2n-2}}.
\ee
Then there is a positive constant $C$ such that the sequence $\{\lm_k\}$ lies in the interval
\[
C\|K\|_{L^n(\bfR^n)}^{-\frac{n-2}{2n-2}}\left(\|K\|_{L^1(\bfR^n)}+\|K\|_{L^n(\bfR^n)}\right)^{-\frac n{2n-2}}\leq\lambda_{k+1} \leq \lambda_0=\Phi(v_0)^{-1}\left(1-\frac{\beta}{\beta_0}\right)^{-1}~.\]
\end{corollary}
\begin{proof} From (\ref{joeleq2}) and Lemma \ref{joellem1}, we have
\bea
1 &\geq& \int_{\bfR^n}(\nabla v_{k+1}\cdot \nabla v_k+\theta_{k+1}v_{k+1}v_k)\, \dd x\nn\\
&=& \lambda_{k+1}\left( \Phi(v_k)-\beta \int_{\bfR^n}K(v_k+\beta)^{\frac{n}{n-2}}\,\dd x\right)\nn\\
 &\geq&
\lambda_{k+1}\left(\Phi(v_k)-\beta \left(\int_{\bfR^n}K\,\dd x\right)^{\frac{n-2}{2n-2}}\Phi(v_k)^{\frac{n}{2n-2}}\right)\nn\\
&\geq& \lambda_{k+1} \Phi(v_0)\left(1-\frac{\beta}{\beta_0}\right).\nn
\eea

Similarly, in view of $v_{k+1}\in \mathcal{A}_{k+1}$, the inequality (\ref{joelGN}), and the bound (\ref{joelPhi}), we also have
\bea
1&=&\lambda_{k+1}\int_{\bfR^n}Kv_{k+1}(v_k+\beta)^{\frac{n}{n-2}}\,\dd x \nn\\
&\leq& \lambda_{k+1}
\|K\|_{L^n(\bfR^n)}^{\frac{n-2}{2n-2}}\left(\int _{\bfR^n}v_{k+1}^{\frac{2n}{n-2}}\,\dd x\right)^{\frac{n-2}{2n}}\Phi(v_k)^{\frac{n}{2n-2} } \nn\\
&\leq& 
 \lambda_{k+1}C \|K\|_{L^n(\bfR^n)}^{\frac{n-2}{2n-2}}\left(\|K\|_{L^1(\bfR^n)}+\|K\|_{L^n(\bfR^n)}\right)^{\frac n{2n-2}},   \nn
\eea
for a uniform constant $C>0$.
\end{proof}

In order to prevent the sequence $\{\lm_k\}$ from diverging to infinity or trivializing at zero, from now on, we assume $\beta$ lies in the interval (\ref{joelbeta})
as stated in Corollary \ref{joelcor1}.\\

\begin{corollary} \label{joelcor2} $\frac12 ||v_{k+1}-v_k||_{k+1}^2 \leq \frac{n-2}{2n-2}\lambda_{k+1} (\Phi(v_{k+1})-\Phi(v_k)))$.
\end{corollary}
\begin{proof}  We have
\bea
\frac12 ||v_{k+1}-v_k||_{k+1}^2 &=&(v_{k+1}, v_{k+1}-v_k)_{k+1}  +
\frac12(||v_k||_{k+1}^2-||v_{k+1}||_{k+1}^2)\nn\\
&\leq& (v_{k+1}, v_{k+1}-v_k)_{k+1} \nn\\
&\leq&   \frac{n-2}{2n-2}\lambda_{k+1} (\Phi(v_{k+1})-\Phi(v_k))),\nn
\eea
by (\ref{joeleq3})  and Corollary \ref{joelcor1} since $\|v_k\|_{k+1}^2\leq \|v_k\|^2_k=1=\|v_{k+1}\|_{k+1}^2 $.
\end{proof}

We next show the sequence $\{v_k\}$ is uniformly bounded and obtain some (non-optimal) decay estimates.

\begin{proposition} \label{joelprop1} There is a uniform constant $C$  such that for $k \geq 1$,\\
 (i) $ \|v_k\|_{L^{\infty}(\bfR^n)} \leq C$,\\
(ii)  $v_k(y) \leq C(1+|y|)^{-a}\,\, $ where $a=\min{(\frac{n-2}2,\frac{(n-2)}{n}l-\varepsilon)},\, 
(0<\varepsilon< \frac{(n-2)}{n}l) $ if $\beta>0$, \\
(iii) $v_k(y) \leq C(1+|y|)^{-\frac{n-2}2 }  $  if $\beta=0$. \\
\begin{proof} Fix a ball $B_R=B_R(y)$ (of radius $R>0$ and centered at $y\in\bfR^n$) in $\bfR^n$. Then, by standard local elliptic estimates, we have
\bea \label{joeleq3.5}\sup_{B_{\frac{R}2}}v_{k+1} &\leq& C\left( R^{-\frac{n}p}\|v_{k+1}\|_{L^p(B_R)}+R^{2-\frac{n}q}\|K(v_k+\beta)^{\frac{n}{n-2}}\|_{L^q(B_R)}\right),\\
&& p>0,\quad q>\frac{n}2.\nn
\eea

We first observe that $v_k(x)\to0$ as $|x|\to\infty$ for all $k\geq1$. In fact, since $K$ is bounded and satisfies (\ref{KL1}), we see that $K(v_0+\beta)^{\frac n{n-2}}\in L^p(\bfR^n)$
for any $p>1$. For $p>n$, elliptic theory applied to (\ref{joeleq1}) implies $v_1\in W^{1,p}(\bfR^n)$. Thus $v_1(x)\to0$ as $|x|\to\infty$. Iterating this argument, we see that $v_{k+1}(x)\to0$
as $|x|\to\infty$ for $k=0,1,2,\cdots$.

To prove (i), let $R=1, \, p=\frac{2n}{n-2}$ and choose $y$ such that $\sup{v_{k+1}}=v_{k+1}(y)$.  Then 
applying (\ref{joelGN}) and $v_k\in \mathcal{A}_k$ ($k\geq1$), we obtain
\be \label{joeleq4}
\sup{v_{k+1}} \leq C\left(1+\varepsilon\left(\int_{B_1} (v_k)^{q+\frac{2n}{n-2}}\,\dd x\right)^{\frac1q}+C(\varepsilon)\right),
\ee
for $\frac{n}2< q< n$ (for example $q=\frac{3n}4$).  Therefore, we find from
(\ref{joeleq4}) and (\ref{joelGN}) the bound
\bea \label{joeleq5}
\sup{v_{k+1}} &\leq& C\left(1+\varepsilon (\sup{v_k})\left(\int_{B_1} v_k^{\frac{2n}{n-2}}\,\dd x\right)^{\frac1q}+C(\varepsilon)\right)
\nn\\
& \leq& C(1+ \varepsilon \sup{v_k} +C(\varepsilon)).
\eea
Now fix $\varepsilon=\frac1{2C}$ to obtain
\be \label{joeleq6} \sup{v_{k+1}} \leq C+\frac12 \sup{v_k}
\ee
for a uniform constant $C$. Iterating, we find  
\be \label{joelconstant}
\sup{v_{k+1}} \leq 2C+ 2^{-(k+1)} \sup{v_0}, 
\ee
proving (i). 

To prove (ii),  let  $2R=|y|, \, p=\frac{2n}{n-2}$. Then, from (\ref{joeleq3.5}), part (i), and the boundedness
of $\Phi(v_k)$, we have
\bea \label{joeleq7} 
v_{k+1}(y) &\leq& C\left(|y|^{-(\frac{n-2}2)}+|y|^{(2-\frac{n}q-\frac{q-1}q l)}\right)\nn\\
&=&C\left(|y|^{-(\frac{n-2}2)}+|y|^{-(l-2+\frac{n-l}q )}\right).
\eea

If $l \geq n$, we let $q=\infty$ and the result follows immediately. If $n>\ell>\frac{n}2$,  then 
$l-2+\frac{n-l}q>\frac{n-2}2$ for $q$ sufficiently close to $\frac{n}2$ and the result follows. If 
$1<l \leq \frac{n}2$, choose $\frac1q=\frac2n-\frac{\varepsilon}{n-l}$, completing the proof of (ii).

Finally if $\beta=0$, we observe that the decay estimates of part ii. are still valid and if we have shown that $v_k(y)\leq C(1+|y|)^{-a},\, a<\frac{n-2}2$, then by the argument of part (ii), we find
\[  v_{k+1}(y) \leq C\left(|y|^{-(\frac{n-2}2)}+|y|^{-(l-2+\frac{n-l}q+a \frac{n}{n-2} )}\right), \]
so that $v_{k+1}\leq C(1+|y|)^{-b}$ with $b=\frac{n}{n-2}a +l(\frac{n-2}n-\varepsilon)>\frac{n}{n-2}a$.
Hence in a finite number of steps we arrive at the decay rate $\frac{n-2}2$ for a (different) uniform constant $C$.
\end{proof}
\end{proposition}

\begin{remark} \label{joelrem1} The dependence of the constant $C$ on $v_0$ comes from the sup norm
estimate of part (i). In particular, from (\ref{joelconstant}), we see that this dependence  will disappear in the limit
as $k$ tends to infinity. Also if we only assume $\int_{\bfR^n} |\nabla v_0|^2 dx \leq 1$ (instead of $\|v_0\|_0
\leq 1$), then,
from (\ref{joeleq3}) and (\ref{joeleq1.8}) with $k=0$, we have
\bea \frac{n-2}{2n-2} \lambda_1(\Phi(v_1)-\Phi(v_0)) &\geq&
\frac12\left(1-\int_{\bfR^n} |\nabla v_0|^2 \,\dd x-\theta_1 \|v_0\|_{L^2(\bfR^n)}^2\right)\nn\\
&\geq& -\frac12  \theta_0 \|v_0\|_{L^2(\bfR^n)}^2. \label{joelnew}
\eea
Hence, in view of (\ref{joelnew}) and Lemma \ref{joellem1}, we have
\be \label{joeleq10}
\Phi(v_{k+1}) \geq \Phi(v_1) \geq \Phi(v_0)-C\theta_0 \|v_0\|_{L^2(\bfR^n)}^2.
\ee
This will be used in the proof of Theorem \ref{joelth2} below.
\end{remark}

We are now ready to state and prove our main existence theorem of this section.

\begin{theorem} \label{joelth1}The sequence $\{v_k\}$ has a subsequence converging locally uniformly in $C^{2+\alpha}$  to a smooth positive function $v$ such that $u=v+\beta$ is a finite energy solution of 
\begin{eqnarray}
-\Delta u &= &\lambda Ku^{\frac{n}{n-2}} \,\,\mbox{ in } \bfR^n, \label{joelsol1} \\
 u(x)&\to& \beta\,\,\mbox{ as }|x| \to \infty, \label{joelsol2}\\
\int_{\bfR^n} |\nabla u|^2\,\dd x &\leq & 1,\label{joelsol3}
\end{eqnarray} 
where $\lambda>0$. If $\beta=0, \,u=O(|x|^{-(n-2)})$ while if $\beta>0,\, u-\beta=O(|x|^{-a})$ where $a=\min{(\frac{n-2}2,\frac{(n-2)}{n}l-\varepsilon)},\, 
(0<\varepsilon< \frac{(n-2)}{n}l) $. Moreover, all estimates are independent of $\sup{v_0}$ by Remark
 \ref{joelrem1}.
\end{theorem}

\begin{proof}  By Lemma \ref{joellem1} and (\ref{joelPhi}), $\{\Phi(v_k)\}$ converges to a finite limit, say $M$. Since the sequence
 $\{v_k\}$ is uniformly bounded in $\mathcal{H}_K \cap L^{\infty}(\bfR^n)$,
 we can choose subsequences $\{v_{k_i}\},\, \{v_{k_i+1}\}$ which converge weakly in $ \mathcal{H}_K \cap L^{\infty}(\bfR^n)$. By Corollary \ref{joelcor2}, both subsequences  converge weakly to the same limit, say $v$. Using that each $v_{k+1}$ satisfies (\ref{joeleq2}) and $\lm_{k+1}$ stays uniformly bounded from above and below (Corollary \ref{joelcor1}),  Proposition \ref{joelprop1} and standard elliptic regularity estimates give that
 the convergence is locally uniformly in $C^{2+\alpha}(\bfR^n)$ and that $u=v +\beta \in C^{\infty}(\bfR^n)$ satisfies (\ref{joelsol1})--(\ref{joelsol3}) and
 \be
\int_{\bfR^n} K(x)u^{\frac{2n-2}{n-2}}(x)\,\dd x =M,
 \ee
where $\lm>0$ is a limiting point of the sequence $\{\lm_k\}$.
 The decay properties of $u=v +\beta$ follow from  Proposition \ref{joelprop1} if $\beta>0$ and  using a bootstrap argument as in part (iii) of  Proposition \ref{joelprop1} in case $\beta=0$.
\end{proof}

We next use Theorem \ref{joelth1} to solve the global constrained variational problem associated with
equations (\ref{joelsol1}), (\ref{joelsol2}). Consider the variational problem
\be
M_{\beta}:=\max \left\{\Phi(v)\,\,\big| \, v \in \mathcal{H}_K,\, \int_{\bfR^n}|\nabla v|^2\,\dd x \leq 1\right\}.
\label{joeleq8}
\ee
 
As shown earlier,  $M_{\beta}$ is always well-defined and the question is whether $M_{\beta}$ 
is achieved.

\begin{theorem} \label{joelth2} Let K be a smooth nonnegative function satisfying $K=O(|x|^{-\ell}),\,\ell>1$.
If $\beta>0$, assume (\ref{KL1}) in addition.
 For $\beta$ in the interval (\ref{joelbeta}), the variational problem $M_{\beta}$
has a finite energy solution $v$ satisfying (\ref{joelsol1}),(\ref{joelsol2}),(\ref{joelsol3}) with the decay properties described in Theorem \ref{joelth1}.
\end{theorem}

\begin{proof} Let $\overline{v}^{(m)}$ be a maximizing sequence for $M_{\beta}$, that is 
$\Phi(\overline{v}^{(m)}) \geq M_{\beta}-\frac1m$, $m=1,2,\cdots$. Without loss of generality we may assume
 $\overline{v}^{(m)} \in  C^{\infty}(\bfR^n) \cap H^1(\bfR^n) \cap L^{\infty}(\bfR^n),\, \overline{v}^{(m)} \geq 0,\,
 \Phi(\overline{v}^{(m)}) \geq \frac{M_0}2$ .
Given $m\geq1$, we choose in our iterative scheme $\theta_0=(m \|\overline{v}^{(m)}\|_{L^2(\bfR^n)}^2)^{-1},\,
v^{(m)}_0=\overline{v}^{(m)}$ and construct the sequence $\{v^{(m)}_k\}$ converging to the solution $v^{(m)}$ (say) of Theorem \ref{joelth1}. Then from Lemma \ref{joellem1}  (see (\ref{joeleq10}) and Remark \ref{joelrem1}),
\bea \label{joeleq9}
\Phi(v^{(m)}) &\geq& \Phi(v^{(m)}_1) \geq \Phi(\overline{v}^{(m)})-C\theta_0\|\overline{v}^{(m)}\|_{L^2(\bfR^n)}^2 \nn\\
&\geq& 
M_{\beta}-\frac{C+1}m.
\eea
Since $\int_{\bfR^n} | \nabla v^{(m)}|^2 \dd x \leq 1$, the sequence $\{v^{(m)}\}$ is also a maximizing sequence
of solutions satisfying
\[ -\Delta v^{(m)} =\lambda^{(m)} K(v^{(m)}+\beta)^{\frac{n}{n-2}}~.\]
By our previous arguments, a subsequence converges  locally uniformly in $C^{2+\alpha}(\bfR^n)$
to a nonnegative solution $v$ of $-\Delta v =\lambda K(v+\beta)^{\frac{n}{n-2}}$ for some $\lm>0$ which is a maximizer of the variational
problem (\ref{joeleq8}).
\end{proof}

A solution $w$ of the original Majumdar--Papapetrou equation (\ref{1}) may then be constructed from a solution $u$ of the nonlinear eigenvalue problem (\ref{joelsol1}) by setting
\be 
w=\lm^{\frac{n-2}2} u.
\ee
\medskip

\section{Nonexistence results}
\setcounter{equation}{0}

For the Majumdar--Papapetrou type boundary value problem
\bea \label{7.1}
\Delta u+K(x)u^{\frac n{n-2}}&=&0,\quad u\geq 0,\quad x\in\bfR^n,\\
\lim_{|x|\to\infty}u(x)&=&\beta,\quad \beta>0,\label{7.2}
\eea
where $K(x)\geq0$ is a smooth function which is not identically zero and satisfies
\be 
\int_{\bfR^n} K(x)\,\dd x<\infty;\quad K(x)=\mbox{O}(|x|^{-\ell}),\quad \ell>2\mbox{ or }\ell>1,
\ee
we have shown that a solution exists when $\beta$ is not too large. Here we note that there may be no solution when $\beta$ is sufficiently large. Without loss of generality, we choose the origin so that
\be \label{7.15}
K(0)=\max_{\bfR^n}K(x)>0
\ee
and set
\be \label{K0}
K_0(r)=\min\{K(x)\,|\,|x|= r, x\in\bfR^n\},\quad r\geq0.
\ee

\begin{lemma}\label{lemma7.1} If the boundary value problem consisting of (\ref{7.1})--(\ref{7.2}) has a solution, then
the  modified  problem,
\be \label{eqv}
\Delta v+K_0(r) v^{\frac n{n-2}}=0,\quad \lim_{|x|\to\infty}v(x)=\beta,\quad v\geq0,
\ee
 has a radially symmetric solution.
\end{lemma}
\begin{proof} Let $u$ be a solution of (\ref{7.1})--(\ref{7.2}). We first show that $u\geq\beta$. In fact, since $u(x)\to\beta$ as $|x|\to\infty$, we see that
$w=\max\{0,\gamma -u\}=(\gamma-u)_+$ is of compact support for any $0<\gamma<\beta$. Multiplying (\ref{7.1}) by $w$ integrating by parts, we have
\be 
-\int_{u\leq \gamma}|\nabla u|^2\,\dd x =\int_{\bfR^n}K(x) u^{\frac n{n-2}} w\,\dd x\geq 0,
\ee
which proves that the set $\{ u<\beta\}$ must be empty as claimed.

On the other hand, using $K_0(r)=K_0(|x|)\leq K(x),x\in\bfR^n$ ($|x|=r$), we have 
\be 
-\Delta u \geq K_0 u^{\frac n{n-2}}.
\ee
In other words, $v^+=u$ is a supersolution of (\ref{eqv}). Taking $v^-\equiv\beta$, we see that $v^-$ is a subsolution of (\ref{eqv}) and $v^-\leq v^+$. We can now construct a radially symmetric solution $v$ , of (\ref{eqv}) satisfying $v^-\leq v\leq v^+$.  This solution is just the minimal positive solution. As before, we solve
\begin{eqnarray*}
-\Delta v_{j+1}&=&K_0 v_j^{\frac{n}{n-2}} \,\, \mbox{in $B_R(0)$}\\
v_{j+1}&=&\beta\,\,\mbox{on $\partial B_R(0)$}\\
v_0 &=&v^-
\end{eqnarray*}
and observe that since the Laplace operator commutes with rotations, the sequence$\{v_j\}$ is radially symmetric by uniqueness. Moreover, 
\[ v^- \leq v_j \leq v_{j+1} <v^+~.\]
Hence $v_{R}(r)=\lim_{j \goto \infty}v_j(r)$ solves
\begin{eqnarray*}
-\Delta v_R&=&K_0 v_R^{\frac{n}{n-2}} \,\, \mbox{in $B_R(0)$}\\
v_R&=&\beta\,\,\mbox{on $\partial B_R(0)$}
\end{eqnarray*}
Finally, in the limit  as $R$ tends to infinity, we obtain a radially symmetric solution, $v$, of (\ref{eqv}) satisfying $v^-\leq v\leq v^+$.
\end{proof}

\begin{proposition} \label{bound} Let $v(r)$ be a radially symmetric solution of the boundary value problem (\ref{eqv}) where $K_0(0)>0$. Then v is apriori bounded by a constant M (independent of $\beta$).
\end{proposition}
\begin{proof}Integrating (\ref{eqv})
over the interval $(0,r)$, we see that $v=v(r)$ decreases. Hence
\be 
v(0)=\max\{v(r)\,|\,r\geq0\}
\ee
 If the proposition is false, 
then we may find a sequence of radially symmetric solutions of (\ref{eqv}), say $\{v_j\}$, satisfying
\be 
M_j\equiv v_j(0)\to \infty \quad j=1,2,\cdots.
\ee
Now  for fixed $j=1,2,\cdots$,  define the rescaled function
\be 
w_j(x)=\frac1{M_j} v_j \left(\frac {x}{M_j^{\frac1{n-2}}}\right)\equiv \frac1{M_j}v_j(\tilde{x}).
\ee
Then $0\leq w_j\leq 1=w_j(0)$ and
\be\label{wj}
-\Delta w_j = M_j^{-\frac n{n-2}}\tilde{\Delta} v_j= M_j^{-\frac n{n-2}}K_0(\tilde{x}) v_j^{\frac n{n-2}}(\tilde{x})
=K_0(\tilde{x}) w_j^{\frac n{n-2}}.
\ee
where $\tilde{\Delta}$ denotes the Laplace operator with respect to the variable $\tilde{x}$.

Using elliptic theory and passing to a subsequence if necessary, we see that the sequence $\{w_j\}$ converges (locally uniformly in $C^{2+\alpha}$) to a $C^2$ function $w$ on $\bfR^n$ with  $w(0)=1$,
which satisfies
\be \label{eqw}
\Delta w+ K_0(0) w^{\frac n{n-2}}=0.
\ee
On the other hand, however, by Theorem 1.1 in \cite{GS}, we have $w\equiv0$, which gives us a contradiction.
\end{proof}

\begin{corollary} \label{beta} Suppose that the function $K$ satisfies (\ref{7.15}).
If the constant $\beta>0$ in (\ref{7.2}) is sufficiently large, the boundary value problem consisting of (\ref{7.1}) and (\ref{7.2}) will have no solution. 
\end{corollary}
\begin{proof} 
If (\ref{7.1})--(\ref{7.2}) has a solution, then (\ref{eqv}) also has a radially symmetric solution as described in Lemma \ref{lemma7.1}.  Hence by Proposition \ref{bound},  $\beta < M$.
\end{proof}

\begin{theorem} \label{interval} There exists $\beta_0>0$ so that 
there is a (minimal) solution of the boundary value problem  (\ref{7.1}) and (\ref{7.2}) for any $0\leq \beta <\beta_0$ and no solution for $\beta>\beta_0$. If $K(x)$ is radially symmetric about the origin with $K(0)>0$,  there is also a solution for $\beta_0$.
\end{theorem} 
\begin{proof}
Define $\beta_0$ be the supremum of the asymptotic values $\beta>0$ such that the boundary value problem  (\ref{7.1}) and (\ref{7.2}) has a solution. As we have shown in the proof of 
Theorem \ref{theorem1}, there exists a minimal positive solution $ u^{(\beta)}(x)$ for $0\leq \beta 
<\beta_0$. Now assume that $K(x)=K_0(|x|)$ is radially symmetric about the origin with $K_0(0)>0$. Then
by Proposition \ref{bound}, the minimal positive solutions $u^{(\beta)}(r)$ with asymptotic value $\beta$ are all uniformly bounded independent of $\beta$.  Hence $u^{(\beta_0)}(r)=\lim_{\beta \to \beta_0}u^{(\beta)}(r)$ is a $C^2$ solution of (\ref{eqv}) with asymptotic value $\beta_0$.
\end{proof}
Next we consider a ``shell"-star type solution of the equation (\ref{7.1}). To motivate our discussion, recall that (\ref{7.1}) allows a many-particle solution
of the form \cite{HH}
\be 
u(x)=c+\sum_{j=1}^N \frac{\mu_j}{|x-p_j|^{n-2}},
\ee
where $c,\mu_1,\cdots,\mu_N$ are positive constants. Such a solution determines the metric induced from a system of $N$ extremely charged particles of masses
$\mu_1,\cdots,\mu_N$ located at the points $p_1,\cdots,p_N\in\bfR^n$ so that the mass density is expressed as a sum of the Dirac distributions concentrated at
$p_1,\cdots,p_N$. In particular, $u(x)\to\infty$ as $x\to p_j$ ($j=1,\cdots,N$).

Similar, an idealized ``shell-star" type solution \cite{GursesH} represents a situation in which
\be 
K(x)=K_0(x)\delta(F(x)),
\ee
where $S=\{x\in\bfR^n\,|\, F(x)=0\}$ is the shell which can be realized as the boundary surface of a bounded domain $\Om$, i.e., $S=\pa\Om$,  and $K_0(x)\geq0$.
Suggested by the afore-discussed many-particle solutions, we may imagine that 
$\Om=\Om_1 \cup \ldots \cup \Om_k$ is the finite union of simply connected compact domains 
and look for a solution of (\ref{7.1})--(\ref{7.2}) in $\bfR^n \setminus \Om$ satisfying the exterior
boundary condition
\be \label{7.16}
u(x)\to\infty,\quad x\to S=\pa\Om.
\ee

However, one easily sees that this is hopeless. In fact,
\begin{lemma} \label{shell}
There is no function u that is  superharmonic ($\Delta u \leq 0$)  and bounded below in $\bfR^n \setminus \Om$ and satisfies
(\ref{7.16}).
\end{lemma}
\begin{proof} Choose the origin to lie in the interior of $\Om$.
By adding a large constant to $u$, we may assume $u \geq \gamma >0$ in $\bfR^n \setminus \Om$.  Then by the (weak) maximum principle, $u \geq \frac{C}{|x|^{n-2}} $ in $\bfR^n \setminus \Om$
 for any positive $C>0$. This is clearly impossible.
\end{proof} 

\section{Conclusions}
\setcounter{equation}{0}

In this paper, we have studied the Majumdar--Papapetrou equation describing an arbitrary continuously distributed extremely charged cosmological dust 
in static equilibrium governed by
the coupled Einstein and Maxwell equations in a general $(n+1)$-dimensional spacetime and established several existence theorems for the solutions of the equation
assuming that the ADM mass is finite. Among the results are the following.

(i) Assume that the mass or charge density vanishes at infinity at a rate no slower than $\mbox{O}(|x|^{-\ell})$ for some positive number $\ell$. When $\ell>2$, a positive solutions approaching a
positive asymptotic value at infinity can be constructed by a sub- and supersolution method. When the condition is relaxed to $\ell>1$, a positive solution with the same properties can be
obtained by an energy method. The obtained solution and the structure of the equation allow us to establish the existence of a continuous monotone family of solutions labeled
by their
positive asymptotic values at infinity which are necessary for achieving asymptotic flatness.

(ii) Specific asymptotic rates of a solution obtained above have been established under various decay assumptions on the mass or charge density. In particular,
if $\ell=n$ and the first derivatives of the mass or charge density decays like $\mbox{O}(|x|^{-(n+1)})$, then the conformal metric of a solution obeys all the  decay rate
assumptions in the classical definition of asymptotic flatness at infinity.

(iii) The set of possible asymptotic values of the conformal metrics determined by the Majumdar--Papapetrou equation is a bounded interval. In other words, there is a positive number
$\beta_0>0$, such that there is no solution $u$ satisfying the asymptotic condition $u(x)\to\beta$ as $|x|\to\infty$ when $\beta>\beta_0$ and there is a solution when $\beta<\beta_0$.

(iv) There is no such ``shell-star" solution of the type that the solution gives rise to an asymptotic flat space and approaches infinity near the shell which is the boundary of
a bounded domain so that the mass density is concentrated at the shell given by a Dirac distribution.

\end{document}